# Eavesdropping on the "ping-pong" quantum communication protocol

Antoni Wôjcik*

Faculty of Physics, Adam Mickiewicz University,

Umultowska 85, 61-614 Poznań, Poland

Abstract

The proposed eavesdropping scheme reveals that the quantum communication protocol recently presented by Bostr÷m and Felbinger [Phys. Rev. Lett. 89, 187902 (2002)] is not secure as far as quantum channel losses are taken into account.

After the pioneering work of Bennett and Brassard published in 1984 [1] a variety of quantum secret communication protocols have been proposed (for a review see [2]). Although there are differences among particular protocols, almost all of them realize the following scenario. First, two strings of classical bits are generated by two legitimate users (Alice and Bob) in some procedure involving transmission through a quantum channel. Then, with the use of a public channel (classical, unjammable channel), each bit string is divided into two parts – verification string and key. The public statistical analysis of verification strings allows Alice and Bob to bound the amount of mutual information between them and a potential eavesdropper (Eve). If this amount of information is too high, the key has to be thrown away. In the other case the procedure of error correction and privacy amplification (also performed with the use of public channel) leads to the final key on which Eve's information is negligible. Let us emphasize here two properties of the above presented general scheme. First, both the verification string and the key are generated by essentially the same procedure. Secondly, this scheme ensures generation of a random key only.

Recently, however, quite a different quantum cryptographic protocol has been proposed by Boström and Felbinger [3]. Their the so-called "ping-pong" protocol allows generation of a deterministic key or even direct secret communication. This improvement is obtained via random switching between two distinct communication modes - message mode and control mode. The key is generated in the message mode, while the eavesdropping is detected in the control mode. The only parameter which has to be analyzed in order to detect the eavesdropper is the correlation of bits generated in the control mode. The established key is believed to be insecure if and only if the results of measurements performed in the control mode coincide. The protocol have been claimed to be secure and experimentally feasible.

The security of the "ping-pong" protocol can be, however, impaired as far as the realistic, not-negligible-distance implementations of this protocol are considered. Which protocol can be considered as practical and secure was specified by Brassard, Lýtkenhaus, Mor and Sanders : "In order to be practical and secure, a quantum key distribution scheme must be based on existing - or nearly existing - technology, but its security must be guaranteed against an eavesdropper with unlimited computing power whose technology is limited only by the laws of quantum mechanics" [4]. The aim of our paper is to present an eavesdropping scheme which allows Eve to obtain some information about the key without any chance of being detected by a procedure proposed by Boström and Felbinger [3]. The scheme works provided that quantum channel losses are not too low, even if perfect photon sources and perfect detectors are used by Alice and Bob. The superiority of Eve over current technology is restricted to the possibility of near lossless photon transmission and performance of two-photon CNOT gate on polarization qubits. Our scheme considers the opportunity of eavesdropping arising due to a separation of two procedures, namely the verification procedure and the key generation. Note that in the "ping-pong" protocol Eve knows which mode (control or message) was chosen by Alice at the time when she could still manipulate the travel photon. On the other hand, we have to confess that an attack can be easily detected if the traditional form of verification involving some subset of the key (e.g. QBER estimation) is performed.

Let us start with the brief description of the "ping-pong" protocol of Boström and Felbinger [3] (see Fig. 1). Bob prepares two photons in entangled state $|\Psi^+\rangle = (|0\rangle|1\rangle + |1\rangle|0\rangle)/\sqrt{2}$ of the polarization degree of freedom. He stores the first photon (home photon), and sends the second photon (travel photon) through a quantum channel to Alice. After receiving the travel photon Alice randomly switches between control mode and message mode. In the control mode Alice measures the polarization of the travel photon and announces the result in the public channel. After receiving Alice's result Bob also switches to the control mode, i.e. measures the state of the home photon in the same basis and compares results of both measurements, which should be perfectly anticorrelated in the absence of Eve. So, the appearance of identical results is considered to be the evidence of eavesdropping and if it occurs, the transmission is aborted. In the other case, the transmission goes on. In the message mode, on the other hand, Alice decides which value $j \in \{0,1\}$ she will transmit to Bob. She encodes this value with the use of the unitary operation $Z^j$, where $Z = |0\rangle\langle 0| - |1\rangle\langle 1|$, performed on the travel photon. The travel photon is then send back to Bob, who measures the state of both photons in the Bell basis. There are only two possible outcomes of this measurement, namely $|\Psi^+\rangle$ or $|\Psi^-\rangle = (|0\rangle|1\rangle - |1\rangle|0\rangle)/\sqrt{2}$. Note that such a restricted Bell measurement can be easily performed [5]. The above result allows Bob to decode the information send to him by Alice. $|\Psi^+\rangle$ encodes $j=0$, while $|\Psi^-\rangle$ encodes $j=1$.

Eve, of course, has no access to the home photon but can manipulate the travel photon while it goes from Bob to Alice and back from Alice to Bob. It was proved by Boström and Felbinger [3] that the eavesdropping strategy which has zero probability of being detected, does not provide any information about the key to Eve. The proof, however, does not take into account the possible transmission losses. We will now present the effective eavesdropping strategy which never produces the identical results of the measurements performed by Bob and Alice in the control mode. The price which has to be paid by Eve is the creation of additional losses in the transmission from Bob to Alice. These losses can be used to detect eavesdropping in the case of ideal channel. On the other hand, in the realistic case of lossy channel, Eve can replace the original channel by a better one and hide the eavesdropping losses in the channel losses.

The lossy quantum channel is described by a single-photon transmission efficiency $\eta$. In order to explain the construction of our protocol, let us first consider the case of the ideal channel ($\eta=1$). Eve uses two auxiliary spatial modes $x$, $y$ together with a single photon in the state $|0\rangle$. She attacks the quantum channel twice, for the first time during the transmission from Bob to Alice (B-A attack) and for the second time during the transmission from Alice to Bob (A-B attack). The eavesdropping protocol (outlined in Fig. 2) starts with preparing two auxiliary modes $x$, $y$ in the state $|\text{vac}\rangle_x |0\rangle_y$, where $|\text{vac}\rangle$ denotes the empty mode. The state of the whole system is thus

$$|\text{initial}\rangle = |\Psi^+\rangle_{ht} |\text{vac}\rangle_x |0\rangle_y \tag{1}$$

when the B-A attack takes place. This attack consist of performing unitary operation Q on three spatial modes t, x, y, where t denotes the travel photon mode. The operation Q defined as

$$Q_{txy} = SWAP_{tx} \, CPBS_{txy} \, H_y \tag{2}$$

is composed of the Hadamard gate, SWAP gate and the three-mode gate which we call the controlled polarizing beam splitter (CPBS). The possible construction of CPBS (presented in Fig. 3) uses CNOT gates and a polarizing beam splitter which is assumed to transmit (reflect) photons in the state $|0\rangle$ ($|1\rangle$). The CPBS, when acting on the relevant states, performs the following transformation

$$\left.\begin{array}{l} |0\rangle|vac\rangle|0\rangle \\ |0\rangle|vac\rangle|1\rangle \\ |1\rangle|vac\rangle|0\rangle \\ |1\rangle|vac\rangle|1\rangle \end{array}\right\} \xrightarrow{CPBS} \left\{\begin{array}{l} |0\rangle|0\rangle|vac\rangle \\ |0\rangle|vac\rangle|1\rangle \\ |1\rangle|vac\rangle|0\rangle \\ |1\rangle|1\rangle|vac\rangle \end{array}\right. . \tag{3}$$

The B-A attack transforms the whole system to the state $|B-A\rangle = Q_{txy}|\Psi^+\rangle_{ht}|vac\rangle_x|0\rangle_y$ of the form

$$|B-A\rangle = \frac{1}{2}|0\rangle_h \left(|vac\rangle_t|1\rangle_x|0\rangle_y + |1\rangle_t|1\rangle_x|vac\rangle_y\right) + \frac{1}{2}|1\rangle_h \left(|vac\rangle_t|0\rangle_x|1\rangle_y + |0\rangle_t|0\rangle_x|vac\rangle_y\right). \tag{4}$$

One sees that the operation Q first transforms the auxiliary photon to a superposition of the polarization states $(|0\rangle + |1\rangle)/\sqrt{2}$ and then sends to Alice one element of this superposition conditionally on the state of the travel photon. The travel photon is stored by Eve. Suppose now that Alice switches to the control mode and measures the state of the mode t. Eq. 4 tell us that with a probability 1/2 Alice detects no photon. However, if the photon is detected, its state is perfectly anticorrelated with the state of the home photon. So, the probability of eavesdropping detection based on the correlation observation equals zero. (The eavesdropping

can be, however, still detected by the observation of the losses.) Let us now analyze the performance of the protocol in the case of Alice operating in the message mode. After Alice performs $Z_t^j$ operation an A-B attack takes place. The A-B attack consist of performing an operation $Q_{txy}^{-1}$. After this attack the state of the system $|A-B\rangle = Q_{txy}^{-1} Z_t^j |B-A\rangle$ is

$$|A-B\rangle = \frac{1}{\sqrt{2}}\left(|0\rangle_h |1\rangle_t |j\rangle_y + |1\rangle_h |0\rangle_t |0\rangle_y\right)|vac\rangle_x .  \quad (5)$$

The final step of the eavesdropping protocol is a measurement of polarization performed on the y- photon. The result of this measurement will be denoted $k$, while the result of Bob's measurement will be denoted by $m = 0\,(1)$ according to the $|\Psi^+\rangle\,(|\Psi^-\rangle)$ result. Let us rewrite Eq. 5 in a more convenient form

$$|A-B\rangle = \frac{1}{2}\left(|\Psi^+\rangle_{ht}|j\rangle_y + |\Psi^-\rangle_{ht}|j\rangle_y + |\Psi^+\rangle_{ht}|0\rangle_y - |\Psi^-\rangle_{ht}|0\rangle_y\right) . \quad (6)$$

Eq. 6 allows us to write the probabilities $p_{jkm}$ of possible measurement's outputs for a given value of $j$. The only nonzero probabilities are

$$\begin{aligned} p_{000} &= 1/2 \\ p_{100} &= p_{101} = p_{110} = p_{111} = 1/8 \end{aligned} \quad (7)$$

Assuming that Alice sends both values of $j$ with the same probability the mutual information between any two parties can be calculated.

$$\begin{aligned} I_{AE} &= I_{AB} = \frac{3}{4}\log_2 \frac{4}{3} \approx 0.311 \\ I_{BE} &= 1 + \frac{5}{8}\log_2 5 - \frac{3}{2}\log_2 3 \approx 0.074 \end{aligned} \quad (8)$$

One sees that mutual information between Eve and Alice equals to the mutual information between Bob and Alice. One can also see that the eavesdropping induces QBER (given by $\sum_k (p_{0k1} + p_{1k0})$) at the level of $1/4$. Note that the scheme is not symmetric in that sense that both the information obtained by Eve and the QBER depend on the value of the bit generated by Alice. Eve can remove this asymmetry by performing with probability of $1/2$ the additional unitary operation $S_{ty}$ just after the operation $Q_{txy}^{-1}$ during the A-B attack. The operation $S_{ty}$ is composed of $Z$, negation $X$ and controlled negation CNOT, namely

$$S_{ty} = X_t \, Z_t \, CNOT_{ty} \, X_t . \tag{9}$$

If the $S_{ty}$ is performed the final state of the system $|A-B\rangle^{(S)} = S_{ty} |A-B\rangle$ is

$$|A-B\rangle^{(S)} = \frac{1}{2}\left(|\Psi^+\rangle_{ht}|j\rangle_y + |\Psi^-\rangle_{ht}|j\rangle_y - |\Psi^+\rangle_{ht}|0\rangle_y + |\Psi^-\rangle_{ht}|0\rangle_y \right) . \tag{10}$$

The symmetrization procedure does not touch the QBER, however, it reduces the mutual information between Alice and Bob to the value

$$I_{AB} = \frac{3}{4} \log_2 3 - 1 \approx 0.189 . \tag{11}$$

So far, we have presented eavesdropping protocol which produces losses and errors but indeed does not produce the correlated results in the control mode. The losses induced by Eve can be, however, hidden in the channel losses. Suppose that Alice and Bob use a quantum channel of $\eta$ not exceeding $50\%$. Typical values of $\eta$ for long-distance experimental quantum key distribution well fit this range [6 - 8]. Eve can replace the original quantum channel by a better one to double its transmission. If the transmission efficiency of Eve's channel is $2\eta$ then the total efficiency (taking into account both channel and eavesdropping losses) seen by Alice equals the efficiency of the original channel i.e. $\eta$. On the other hand, the efficiency of the transmission Bob - Alice - Bob in the message mode should be $\eta^2$ not $4\eta^2$. So Eve has to filter out $75\%$ of the photons reaching the Bob in the message mode. In this way the information about the eavesdropping is completely erased from the data generated in the control mode. If the efficiency of the original channel $\eta$ exceeds $50\%$, the undetectable eavesdropping is still possible, however, mutual information $I_{AE}$ cannot reach the value given in Eq. 10. In this case Eve has to replace the original channel by the ideal one and to eavesdrop only the fraction $\mu = 2(1-\eta)$ of the transmitted bits. The values of the mutual information $I_{AE}$ and $I_{AB}$ as functions of $\eta$ are presented in Fig.4. It can be seen that the mutual information between Eve and Alice can exceeds the mutual information between Bob and Alice up to almost 60% transmission efficiency.

Let us now consider how to improve of the "ping-pong" protocol to make it secure. This can be done, e.g., in a traditional way by sacrificing some part of the key in order to estimate QBER. Our scheme produces QBER equal to 25% which should be easily detected as the QBER measured in the long-distance quantum key distribution experiments [6-8] does not exceeds a level of a few percents. There is, however, another way to protect the "ping-pong" protocol against eavesdropping. Note that Eve's action depends on the actual Alice's choice of the communication mode (control or message). If, e.g., she performs the A-B attack in the case of switching to the control mode by Alice it could happen that both Alice and Bob detect the photon in the travel mode. Such a "double" detection of the single travel photon can be used as an additional evidence of Eve's action. Thus, Alice has to delay the announcement of the information about a chosen mode. Bob, on the other hand, apart from measuring of the home photon's polarization has also to check if there is any photon in the travel mode. In this way the detection of eavesdropping based on the analysis restricted to the control mode can be achieved.

In conclusion we have presented undetectable eavesdropping scheme working on the realistic implementation of the "ping-pong" quantum communication protocol. The eavesdropping scheme works if the quantum channel losses cannot be ignored. It exploits the fact that the "ping-pong" protocol is performed with the use of two distinct modes (control and message mode), and moreover, that the information about which one of them is actually used is revealed too soon, i.e. in the time when Eve still has access to the travel photon. We also suggest the way in which the original "ping-pong" protocol can be improved to fulfill the conditions of both practicality and security.

We would like to thank the State Committee for Scientific Research for financial support under grant no. 0 T00A 003 23.

*Email address: antwoj@amu.edu.pl

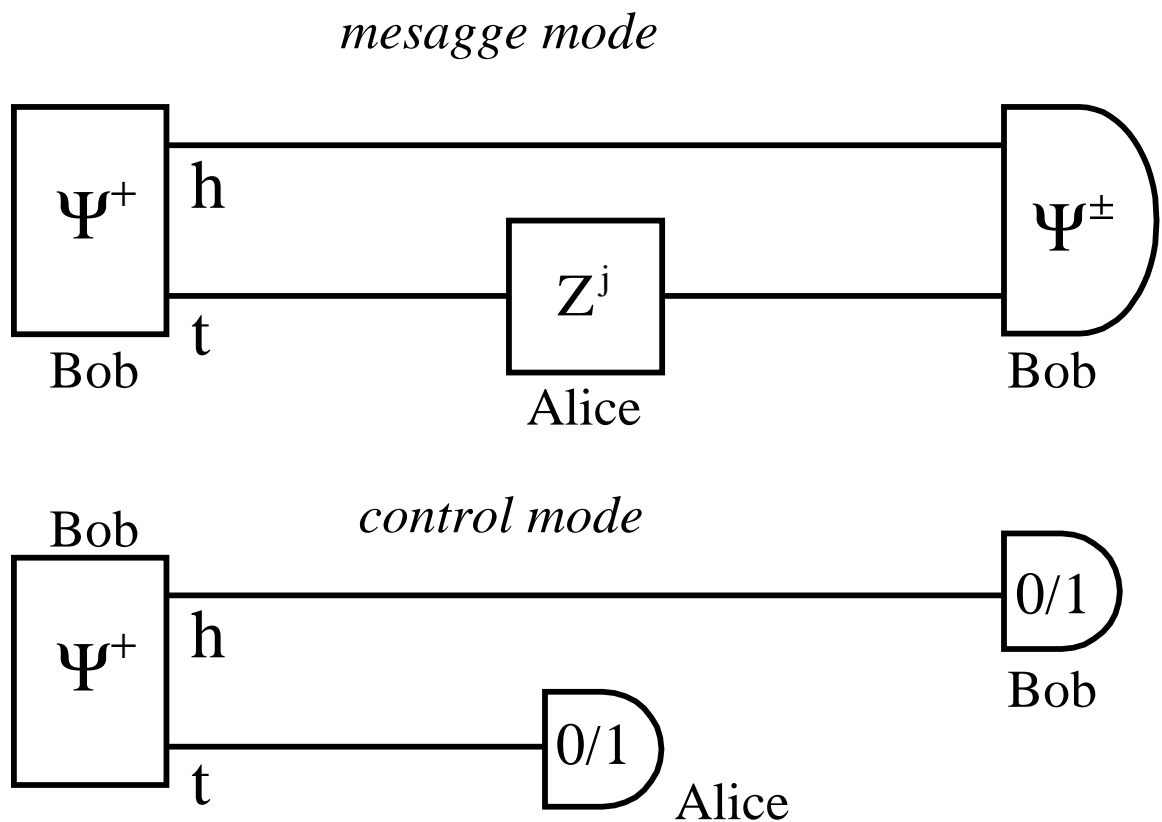

Fig. 1 The message mode and the control mode of the "ping-pong" protocol; h and t denote the home and the travel photon, respectively.

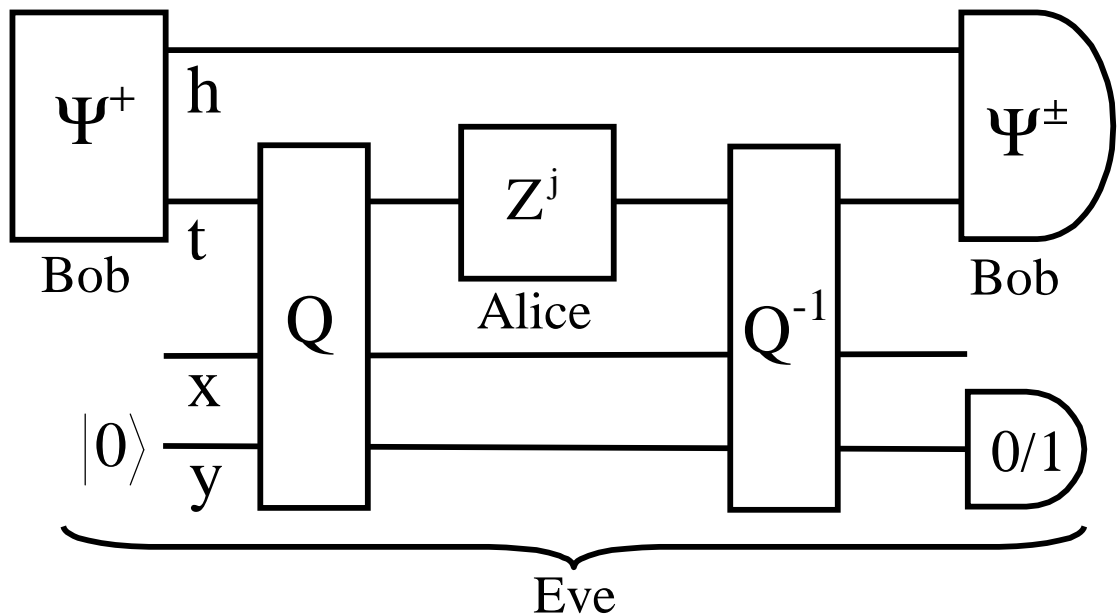

Fig. 2 Eavesdropping on the "ping-pong" protocol.

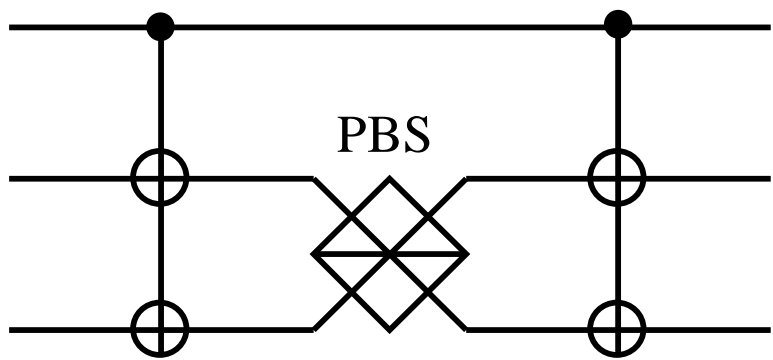

Fig. 3 Controlled polarization beam splitter (CPBS). The polarization beam splitter (PBS) transmits (reflects) photons in the state $|0\rangle$ ($|1\rangle$).

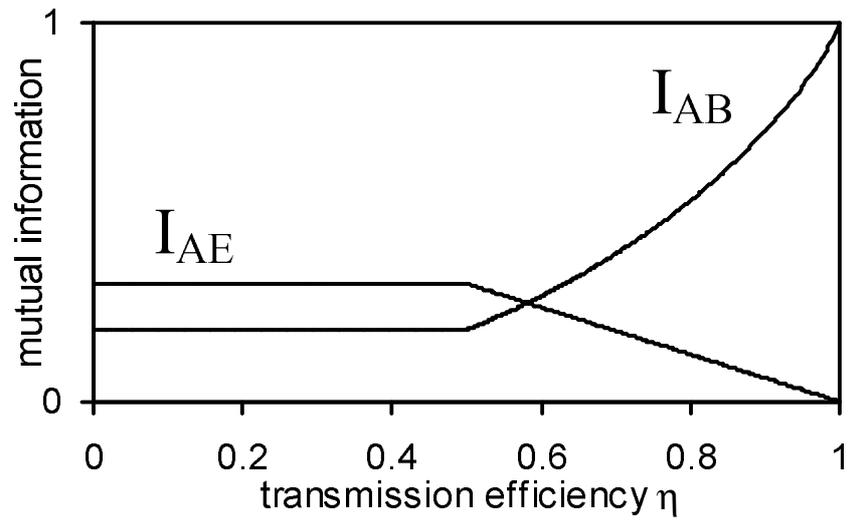

Fig. 4 Mutual information between Eve and Alice $I_{AE}$ and mutual information between Bob and Alice $I_{AB}$ as a function of quantum channel transmission efficiency $\eta$.